\documentstyle[aps,12pt,epsf]{revtex}
\newcommand\pubdate{\today}
\newcommand\hepnumber{hep-ph/0306029}

\def\csumb{Ottawa-Carleton Institute for Physics,
Department of Physics, Carleton University, Ottawa, Canada K1S 5B6
}

\def\Title#1{\begin{center} {\Large\bf #1 } \end{center}}
\def\Author#1{\begin{center}{ \sc #1} \end{center}}
\def\Address#1{\begin{center}{ \it #1} \end{center}}

\newcommand\pubblock{\rightline{\begin{tabular}{l}
         \pubdate\\ \hepnumber \end{tabular}}}
\newenvironment{Abstract}{\begin{quotation}  }{\end{quotation}}

\def\beq{\begin{equation}}
\def\eeq{\end{equation}}
\def\beqn{\begin{eqnarray}}
\def\eeqn{\end{eqnarray}}
\def\bea{\begin{eqnarray}}
\def\eea{\end{eqnarray}}
\def\be{\begin{equation}}
\def\ee{\end{equation}}

\begin{document}
\begin{titlepage}
\pubblock

\Title{ $b\rightarrow s \gamma$ in the Littlest Higgs Model }
\vfill \Author{ Wujun Huo and Shouhua Zhu   } \Address{\csumb}
\vfill
\begin{Abstract}
The inclusive process $b \rightarrow s \gamma$ is studied in the
littlest Higgs model. The contributions arising from new particles
are normally suppressed by a factor of $O(v^2/f^2)$. Due to the
large uncertainties of experimental measurements and theoretical
predictions, the model parameters can escape from the constraints
of present experiments provided $f \ge 1$ TeV.

\end{Abstract}
\vfill


\end{titlepage}

\eject \baselineskip=0.3in

%\begin{narrowtext}

%\section{Introduction}

Little Higgs (LH) models
\cite{Arkani-Hamed:2001nc,Arkani-Hamed:2002pa,Arkani-Hamed:2002qx,Low:2002ws,Arkani-Hamed:2002qy,Schmaltz:2002wx},
as an alternative approach to supersymmetric models, are invented
to stabilize light Higgs boson mass by introducing new gauge
bosons, scalars and quarks. Unlike the case of supersymmetric
models, the cancellation of quadratic divergences are realized
through the same spin particles. The physical picture is that
below TeV scale, the physics can be approximately described by the
standard model (SM), for higher energy scale of order of $O(10)$
TeV, the new particles might emerge. It is obvious that LH is not
the end of the story, ultraviolet completion of the theory must be
explored which is beyond the scope of this paper.

Based on the idea of the LH, a model named ``littlest Higgs model"
\cite{Arkani-Hamed:2002qy} has been constructed and its explicit
interactions have been presented in Ref. \cite{Han:2003wu}. Lots
of phenomenological studies in this model have been performed
\cite{Han:2003wu,Csaki:2002qg,Hewett:2002px,Burdman:2002ns,Dib:2003zj,Han:2003gf,Chang:2003un,Csaki:2003si,Kribs:2003yu}.
In this brief report, we concentrate on the effects of new
particles on the inclusive process $b \rightarrow s \gamma$, which
is known as an ideal
place to study new flavor physics \cite{Hurth}. \\

%\section{Formulas}

In order to demonstrate the new physics effects, we use the
leading order results to estimate the branching ratio for
inclusive process $b \rightarrow s \gamma$, and
\begin{eqnarray}
Br^{LH}(b \rightarrow s \gamma) =
Br^{SM}(b \rightarrow s \gamma) \left( \frac{C^{LH}_{7\gamma}(m_b)}{C^{SM}_{7\gamma}(m_b)}
\right)^2.
\end{eqnarray}
It is well known that the $C_{7\gamma}$ at scale of $m_b$ can be
easily obtained from $C_{7\gamma}$,  $C_{8G}$ and $C_2$ at $m_W$
scale through renormalization group equations
(RGE)\cite{BurasReview}.

The Wilson Coefficients at $m_W$ scale can be generally written as
\begin{eqnarray}
C^{LH}_x(m_W) &=& C^{SM}_{x}(m_W) \left[ 1+\Delta^{LH}_{x}
\right], \label{c7mw}
\end{eqnarray}
where $x$ represents $7\gamma$, $8G$ or $2$, and $\Delta^{LH}$
arises respectively from unitarity violation (V) of CKM matrix in
the SM and the new charged gauge bosons as well as the new fermion
$T$ and new charged Higgs bosons $\Phi^\pm$,
\begin{eqnarray}
\Delta^{LH}_{x} &=&
\Delta^{V}_x+\Delta^{T}_x+\Delta^{W_H}_x+\Delta^{W_HT}_x+\Delta^{\Phi^\pm}_x+
\Delta^{\Phi^\pm T}_x.
\end{eqnarray}
Here
\begin{eqnarray}
C^{SM}_{7\gamma}(m_W) &=& - \frac{A(x_t)}{2} \nonumber  \\
C^{SM}_{8G} (m_W) &=& -\frac{D(x_t)}{2} \nonumber  \\
 C^{SM}_{2}
&= & 1
\end{eqnarray}
with
\begin{eqnarray}
A(x) &=& x \biggr [ \frac{8x^2 +5x -7}{12(x-1)^3} - \frac{(3x^2
-2x)\ln x}{2(x-1)^4} \biggr ] \nonumber  \\
D(x)&=& x\biggr[
\frac{x^2 - 5x -2}{4(x-1)^3} + \frac{3x\ln x}{2(x-1)^4}\biggr ].
\end{eqnarray}
$\Delta_{7\gamma}$ 
can be written to $O(v^2/f^2)$ as 
[$\Delta_{8G}$ can be obtained by replacing $A$ in Eq. (\ref{LHE})
as $D$] 
\begin{eqnarray}
\Delta^{V} &=& -\frac{v^2}{f^2} \left(
c^2 (c^2-s^2)+x_L^2 \right), \nonumber \\
\Delta^{T} &=&
\frac{v^2}{f^2} x_L^2 \frac{A(x_T)}{A(x_t)},  \nonumber \\
\Delta^{W_H} &=& \left[ \left(\frac{c}{s}\right)^2+
\frac{v^2}{f^2} \left( c^2 (c^2-s^2)-\left(\frac{c}{s}\right)^2 x_L^2 \right)
\right] \frac{A(x_{w_H})}{A(x_t)} \frac{m_W^2}{m_{W_H}^2}, \nonumber \\
\Delta^{W_HT} &=& \frac{v^2}{f^2} \left(\frac{c}{s}\right)^2 x_L^2
\frac{A(x_{w_HT})}{A(x_t)}\frac{m_W^2}{m_{W_H}^2},
\nonumber \\
 \Delta^{\Phi^\pm} &=&
 \frac{|\frac{v}{f}-2 s_{+}|^2}{12} \frac{ \left[A(x_{\Phi^\pm})+
6 B(x_{\Phi^\pm}) \right] }{A(x_t)}, \nonumber \\
\Delta^{\Phi^\pm T} &=&  \frac{|\frac{v}{f}-2 s_{+}|^2}{12} \frac{
\frac{\lambda_1^2}{ \lambda_2^2} A(x_{\Phi^\pm T})}{A(x_t)}
\frac{m_t^2}{m_T^2} \label{LHE}
\end{eqnarray}
with $x_t=m_t^2/m_W^2$, $x_T=m_T^2/m_W^2$,
$x_{w_H}=m_t^2/m_{W_H}^2$, $x_{w_HT}=m_T^2/m_{W_H}^2$,
$x_{\Phi^\pm}=m_t^2/m_{\Phi^\pm}^2$, $x_{\Phi^\pm
T}=m_T^2/m_{\Phi^\pm}^2$,
$x_L^2=\frac{\lambda_1^4}{(\lambda_1^2+\lambda_2^2)^2}$ and $s_{+}
\approx 2 v^\prime/v <v/(2 f)$. 
Here, $B$ can be written as
\begin{eqnarray}
B(y) = \left\{
\begin{tabular}{cc}
$\frac{y}{2} \left[ \frac{ \frac{5}{6} y -\frac{1}{2}}{(y-1)^2}-
\frac{y-\frac{2}{3}}{(y-1)^3} \log y \right] $, & {\rm for \
$\Delta^{\Phi^\pm}_{7 \gamma}$ }\\
$\frac{y}{2} \left[ \frac{ \frac{1}{2} y -\frac{3}{2}}{(y-1)^2}+
\frac{1}{(y-1)^3} \log y \right]$, & {\rm for \
$\Delta^{\Phi^\pm}_{8 g}.$}
\end{tabular}
\right.
\end{eqnarray} 
In Eq. (\ref{LHE}), $f$ is the
scale where new physics enters,  $\lambda_{1,2}$ are parameters in
yukawa interactions which give ``raw" masses to SM fermions and
vector-like top quark and are supposed to be the order of unity
\cite{Han:2003wu}, and $c$ and $s$ are $sin$ and $cos$ of the
charged sector mixing angle $\theta$ when the Higgs field breaks
$[SU(2)\otimes U(1)]^2$ into its diagonal subgroup $[SU(2)\otimes
U(1)]_{SM}$. It should be noted that the main contributions come
from the first two terms in Eq. (\ref{LHE}), which are only
suppressed by $O(v^2/f^2)$. And $\Delta^{LH}_{2}$ can be expressed
as
\begin{eqnarray}
\Delta^{LH}_{2} &=& - \frac{v^2}{f^2} c^2 (s^2-s^2)+ \frac{x_{w_H}}{x_t} \frac{c^2}{s^2}.
\end{eqnarray}\\

%\section{Numerical Results}

In the following we present some numerical analysis and adopt the
mass relation of new particles at leading order as
\begin{eqnarray}
\frac{m_{w_H}}{m_w} &=& \sqrt{\frac{1}{s^2 c^2}  \frac{f^2}{v^2} -1} \approx
\frac{1}{s c} \frac{f }{v} \nonumber \\
\frac{m_T}{m_t}     &=& \frac{\lambda_1^2+\lambda_2^2}{\lambda_1 \lambda_2}
\frac{f}{v} \nonumber \\
\frac{m_{\Phi^\pm}}{m_H} &=&
\sqrt{2} \frac{f}{v}
\label{massrelation}
\end{eqnarray}
with $m_H=115$ GeV. Motivations of little Higgs model imply that
the masses of additional Higgs bosons and gauge bosons are order
of TeV. Therefore from Eq. (\ref{massrelation}) we must require
that $\frac{1}{sc}$ can not be too large. In our numerical
calculations we choose $\frac{1}{2} <\tan\theta <10$, which
corresponds to $\frac{1}{sc} <10$. At the same time, we omit the
$s_+$ contribution.

The SM theoretical estimation is, at next-to-leading order\cite{Hurth},
\begin{eqnarray}
Br^{SM}(b \rightarrow s \gamma)=(3.32\pm 0.30) \times 10^{-4}.
\end{eqnarray}
However, because the new physics contributions are only calculated to leading
order, we adopt here the leading-order results in the SM as \cite{Buras:xp},
\begin{eqnarray}
Br^{SM}(b \rightarrow s \gamma)=(2.8\pm 0.8) \times 10^{-4},
\end{eqnarray}
and the experimental measurement is quoted as \cite{Hagiwara:fs}
\begin{eqnarray}
Br(b \rightarrow s \gamma)=(3.3\pm 0.4) \times 10^{-4}.
\end{eqnarray}

We have scanned the parameter space and find that the parameters
can escape the constraints from the experiment measurements
provided $f \ge$ 1 TeV. In order to demonstrate the new physics
effects, in Fig.~1, we show the relative correction $$ \delta
=\frac{{\rm Br}^{LH}-{\rm Br}^{SM}}{{\rm Br}^{SM}}$$ as a function
of $f/v$ with $\lambda_1/\lambda_2=5$. From the figure, it is
obvious that effects arising from new particles in the littlest
Higgs model can change the SM value at a level of a few
percents with $f/v =5 \sim 20$.  \\

To summarize, the contributions to inclusive process $b
\rightarrow s \gamma$ from new particles in the littlest Higgs
model have been studied. The new physics effects are suppressed at
least by a factor of $O(v^2/f^2)$ and can escape the constraints
from $b \rightarrow s \gamma$ for $f \ge $ 1 TeV. We note that the
constraints from $b \rightarrow s \gamma$ is relatively loose due
to the large theoretical and
experimental uncertainties. \\

%\section*{Acknowledgements}
This work was supported in part by the Nature Sciences and Engineering Research Council of Canada.
SZ is also supported in part by Nature Science Foundation of China and acknowledges the hospitality
of University of Wisconsin, as well as the useful conversation with T. Han and L.T. Wang and F. Dalnoki-Veress.
WH also thanks Prof. Xinmin Zhang and Prof. Cai-Dian L\"{u} for the
useful suggestions.
%\end{narrowtext}
%\vspace{-0.7truecm}

\newpage
\begin{figure}
\epsfxsize=12 cm \centerline{\epsffile{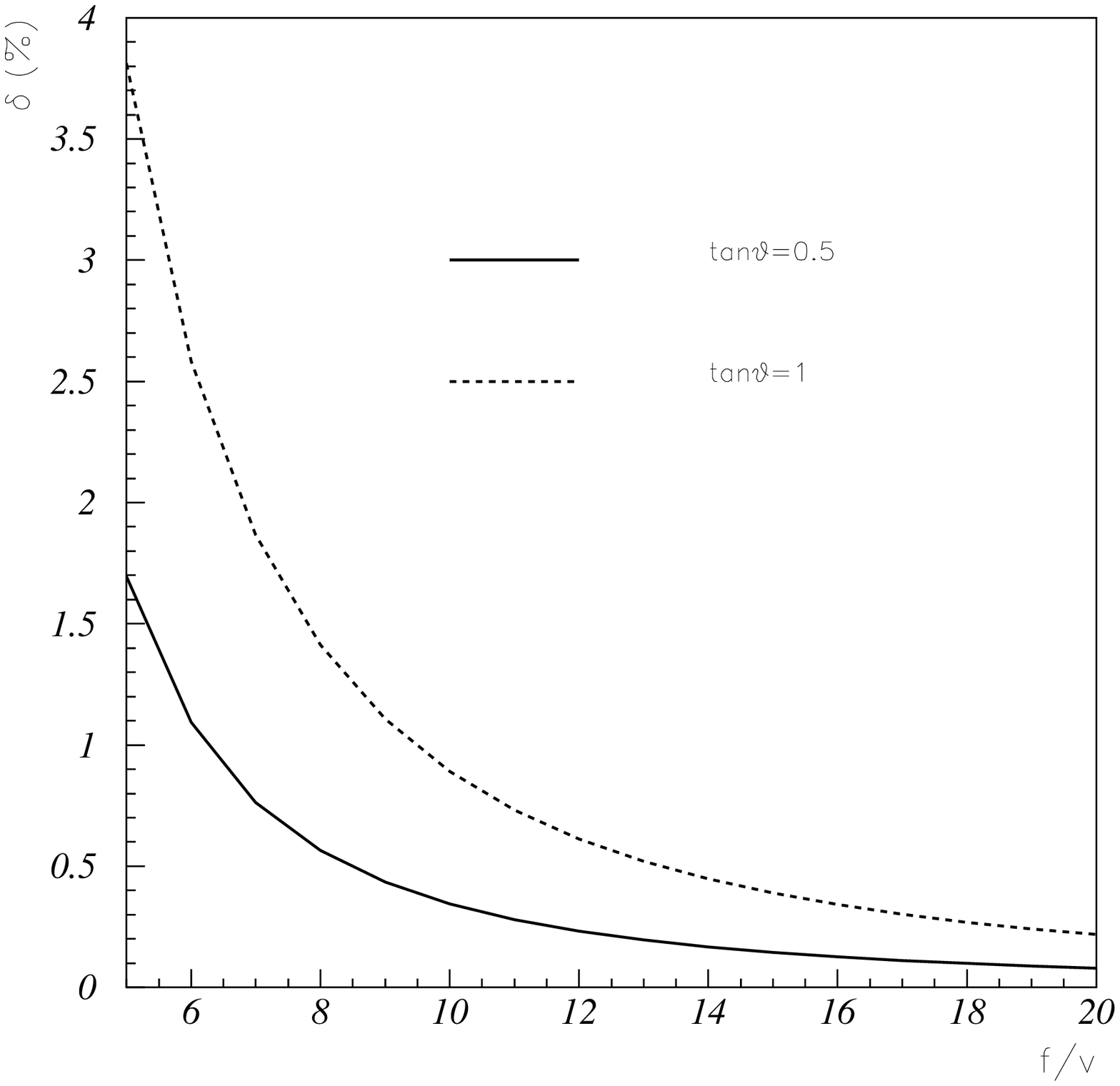}}
%\centerline{\epsfig{file=fig.eps,width=14cm,clip=}} \vspace{20pt}
\caption[]{ The relative correction $\delta=\frac{{\rm
Br}^{LH}-{\rm Br}^{SM}}{{\rm Br}^{SM}}$ as a function of $f/v$
with $\lambda_1/\lambda_2=5$. }
\end{figure}

\end{document}